\documentclass[useAMS,usenatbib]{emulateapj}
\usepackage{graphicx}
\usepackage{epsfig}
\usepackage{natbib}
\usepackage{verbatim}
\usepackage{amssymb,amsmath}
\usepackage{verbatim}

\usepackage{color}

\DeclareMathVersion{bold}

\newcommand{\src}{A2261-BCG}
\newcommand{\eg}{e.\,g.}
\newcommand{\msun}{M_{\odot}}

\newcommand{\asec}{\hbox to 1pt{}\rlap{$^{\prime\prime}$}.\hbox to 2pt{}}
\newcommand{\amin}{\hbox to 1pt{}\rlap{$^{\prime}$}.\hbox to 1pt{}}
\newcommand{\adeg}{\hbox to 1pt{}\rlap{$^{\circ}$}.\hbox to 2pt{}}
\newcommand{\ras}{\hbox to 1pt{}\rlap{$^{s}$}.\hbox to 2pt{}}
\newcommand{\kms}{{\rm km\,s^{-1}}}

\shorttitle{A Search for the \src\ Black Hole}
\shortauthors{Burke-Spolaor et al.}

\begin{document}

\title{A Radio Relic and a Search for the Central Black Hole in the Abell 2261 Brightest Cluster Galaxy}
%\unnecessarysubtitle{Seeking: Recoiling Black Hole in the Abell 2261 Brightest Cluster Galaxy. Finding: Dwarf Galaxies and a Radio Relic.}

\author{Sarah Burke-Spolaor\altaffilmark{1,2,3,4}
Kayhan G\"ultekin\altaffilmark{5},
Marc Postman\altaffilmark{6},
Tod R. Lauer\altaffilmark{7},
Joanna M. Taylor\altaffilmark{6},
T.~Joseph~W.~Lazio\altaffilmark{8},
and Leonidas A. Moustakas\altaffilmark{8}
}
\altaffiltext{1}{Center for Gravitational Waves and Cosmology, West Virginia University, Chestnut Ridge Research Building, Morgantown, WV 26505. Email: sarah.spolaor@mail.wvu.edu}
\altaffiltext{2}{Department of Physics and Astronomy, West Virginia University, Morgantown, WV 26506, USA}
\altaffiltext{3}{National Radio Astronomy Observatory, Socorro, New Mexico 87801, USA}
\altaffiltext{4}{Jansky Fellow for a portion of this work.}
\altaffiltext{5}{University of Michigan, Department of Astronomy,
301E West Hall, 1085 S. University Ave., Ann Arbor, MI 48109}
\altaffiltext{6}{Space Telescope Science Institute, 3700 San Martin Drive,
Baltimore, MD 21218}
\altaffiltext{7}{National Optical Astronomy Observatory, P.O. Box 26732, Tucson, AZ 85726}
\altaffiltext{8}{Jet Propulsion Laboratory, California Institute of Technology, 
4800 Oak Grove Dr., Pasadena CA 91106}

%\maketitle

\begin{abstract}
%{NEED TO REFERENCE https://arxiv.org/abs/1707.05336}
We present VLA images and HST/STIS spectra of sources within the center
of the brightest cluster galaxy (BCG) in Abell 2261.
These observations were obtained to test the hypothesis that its extremely
large, flat core reflects the ejection of its  supermassive black hole.
Spectra of three of the four most luminous ``knots'' embedded in the core were taken to test
whether one may represent stars bound to a displaced massive black hole.
The three knots have radial velocity offsets $(|\Delta V|\lesssim 150
{\rm~\kms})$ from the BCG.
Knots 2 and 3 show kinematics, colors,
and stellar masses consistent with infalling {low-mass} galaxies or larger stripped cluster members.
Large errors in the stellar velocity dispersion of Knot 1, however, mean
that we cannot rule out the hypothesis that it hosts a high-mass black hole.
\src\ has a compact, relic radio-source offset by $6.5$~kpc (projected)
from the optical core's center, but no active radio core that would pinpoint the galaxy's central black hole to a tight 10\,GHz flux limit $<3.6\,\mu$Jy.
Its spectrum and morphology are suggestive of an active galactic nucleus that
switched off {$>48$~Myr ago, with an equipartition condition magnetic field of 15\,$\mu$G.}
These observations are still consistent with the hypothesis that the
nuclear black hole has been ejected from its core, but the critical
task of locating the supermassive black hole or demonstrating that
\src\ lacks one remains to be done.
\end{abstract}

\keywords{galaxies: nuclei --- galaxies: kinematics and dynamics ---
galaxies: jets}

\section{Is the Large Core in \src\ Due To an Ejected Black Hole?}
\label{sec:intro}

It is now understood that supermassive back holes are not only common
to the centers of galaxies, but have masses closely tied to the
properties of their hosts \citep{mag,msig,ferraresemerrittmsig}.
A galaxy and its black hole
are formed and evolve together, each influencing the other.
The existence of ``cores'' in the central distribution of stars
in luminous elliptical galaxies are hypothesized to be a prominent example of just such
a mechanism where the action of the nuclear black hole has shaped the central
structure of the galaxy itself.
Cores are regions over which the central stellar density distribution breaks as the center of the galaxy is approached,
in contrast to the steep envelope density profile of the surrounding galaxy.
They were first seen in high resolution images of elliptical
galaxies obtained from the ground \citep{l85, k85},
and studied extensively in Hubble Space Telescope (HST) images
\citep{f94,l95,laine03,l05}.

\citet{postman12} identified \src\ as a galaxy hosting an exceptionally large core.
The stellar surface brightness distribution has a cusp radius (the
radius at which the local logarithmic slope of the profile is $-1/2$)
of $r_\gamma=3.2$~kpc. Its core is thus twice as big as that in the NGC 6166,
the BCG with the largest core in the \citet{lauer07} compilation,
which incorporates the extensive {\it HST} imaging survey of 
nearby BCGs of \citet{laine03}.  \citet{a85} subsequently identified
an even larger core in A85-BCG, but the core in \src\ remains
a strong outlier in comparison to those of nearly all giant elliptical galaxies studied
so far.

\citet{postman12} further show that the \src\ has a completely flat or
even slightly depressed brightness profile interior to the core.
Most cores generally have singular, if shallow, cusps interior to
the cusp radius \citep{l95,l05}.  A small number of galaxies
have cores with centrally {\it decreasing} brightness profiles,
but they are rare \citep{l02,l05}. \citet{postman12} hypothesized that both the large core
and flat structure could be explained if the nuclear black
hole had been ejected from the center of system.

The mass of the expected nuclear black hole (M$_{\bullet}$) in \src\ 
is {predicted to be between $5.6 \pm 1.0 \times 10^9\,\msun$ and approximately $11 \times 10^9\,\msun$, based on the observed velocity distribution ($\sigma = 387\,{\rm \kms}$, Postman et al.) and the $M_\bullet-\sigma$ relations of \citet{korho13} and \citet{mcconnell2011}, respectively.
%predicted to lie in the range $5.6\pm1.0 \times 10^9 M_{\odot}$ (from the the elliptical galaxy M$_{\bullet}\ -\ \sigma$ relation of \cite{korho13}  and the observed \src\ $\sigma = $ 387 km\,s$^{-1}$ \citep{postman12}) to $\sim 1.1 \times 10^{10} M_{\odot}$ (from the \cite{mcconnell2011} M$_{\bullet} - $ Luminosity relation). 
A somewhat higher upper limit on the  mass of
the \src\ nuclear black hole can be derived from the fundamental-plane of the black hole activity \citep{merloni2003}, 
which estimates M$_{\bullet}$ from a combination of the core radio luminosity and the nuclear X-ray emission.
\cite{hlalar2012} use this approach to obtain an estimate of M$_{\bullet} = 2.0^{+8.0}_{-1.6} \times 10^{10} M_{\odot}$ for
\src. }

\citet{begelman80} suggested that the merger of two galaxies,
each hosting a central supermassive black hole,
would form a binary black hole at the center of the merged system.
The binary might then ``scour'' out a core in the merged system
as central stars would gravitationally interact with the binary,
drawing orbital energy from it and causing them to be ejected from the center.
N-body simulations have demonstrated this
phenomenon directly \citep{ebisuzaki91, makino97, milomerritt01}.
\citet{faber97} offered strong observational support for this scenario,
showing that the most luminous elliptical galaxies nearly always have cores,
and are correlated with slow-rotation and ``boxy'' isophotes in these systems,
while less-luminous or rapidly rotating ellipticals rarely have cores.
Core formation is thus a natural end-point of ``dry mergers'' of two progenitor galaxies.

Although there is circumstantial evidence that cores are made by the hardening of a supermassive black hole binary,
direct observational proof for the existence of such binaries is still lacking.
Other phenomena that come into play during the evolution of the binary
may leave their own imprint on the central structure
of the hosting galaxies, however, and thus strengthen the case if observed to be directly linked to a black hole pair.
Direct observation of SMBH binaries may become possible in coming years by gravitational-wave experiments like pulsar timing arrays \citep{time2detPTA} or LISA \citep{LISA}.
%If the initial formation of the binary is followed by a subsequent merger, a three-body interaction with a third black hole newly introduced into the central potential may cause all holes to be ejected from the core.
%Apart from this, t

The terminal hardening of the binary and the subsequent
coalescence of the two holes into one may generate a strong jet of gravitational
radiation that ejects the coalesced product from the core as well \citep{recoilfx1}.
In either case the sudden removal of the black hole mass would
cause the core to rebound and expand in size.  If the black hole is
not ejected from the galaxy, but remains on a radial orbit that
returns it to the center, energy exchange with the stars in the core
via dynamical friction may also enlarge the core well beyond its initial
size as produced by the hardening of the binary \citep{bk04, m04}.

The completely flat core of \src\ strongly resembles
the large cores formed by ejection of the central black hole \citep{gm}.
The core of \src\ is also slightly displaced from the photo-center of the
surrounding galaxy, suggesting a relatively recent dynamical 
disturbance, while the rest of the galaxy shows no evidence for recent
interaction.  Lastly, there are four compact low-luminosity
``knots'' in close proximity to the core.  The ejected black hole
is predicted to carry a tightly bound compact ``cloak" or hypercompact stellar system (HCSS) of stars
with it, potentially similar to these knots \citep{m09}.

%The brightest cluster galaxy in Abell 2261 (hereafter \src) is a galaxy of extremes; at \mbox{$\sigma=387\pm16\,{\rm km\,s^{-1}}$}, it has a stellar velocity dispersion among the highest values observed in all galaxies. It also is one of the most luminous BCGs known, with a total absolute magnitude --24.70 in the $V$ band. Furthermore, \src\ was recently revealed to have a core with two abnormal features: its core radius, \rcusp, is 3.2\,kpc (0.89$''$), more than two times the largest core known; and its surface brightness profile is flat, consistent with scouring by a supermassive black hole pair with a total mass of $\sim$$10^{10}\,\msun$ \citep{postman12}. Postman et al.\ hypothesized that the most likely scenario for this system is that its supermassive binary underwent a gravitational recoil following coalescence, thus further scouring the core. Their analysis of this system indicated several interesting features, including four compact, non-central optical knots within the core region, and a displacement of the central galactic core within its envelope at 0.7$"$ to the northeast. The knots were not identifiable from previously available data, but it remained a hypothesis that one or several of them contain a supermassive black hole. If so, this would lend strong support for a post-coalescence recoil scenario for \src.

The question that motivates this work is: {\it Where is the supermassive
black hole that should reside in this galaxy?}
In \S\ref{sec:knots} we present new (HST)
spectroscopic observations of the stellar knots within the
core of \src\ to measure their redshifts, stellar velocity dispersions,
and to test the possibility that one of them in fact is hosting
a wayward black hole.
In \S\ref{sec:radio} we present new Jansky Very Large Array (VLA) observations
of the radio source in the core of \src\ to test whether active galactic nucleus activity might mark the black hole.
In \S\ref{sec:disco} we discuss how and if the present observations
have advanced or refuted the ejected black hole hypothesis.

Throughout this paper, we assume H$_{\rm o} = 70$ km s$^{-1}$ Mpc$^{-1}$, $\Omega_\Lambda = 0.7$, and $\Omega_m = 0.3$. 
%With these cosmological parameters, the projected physical scale at the redshift of A2261 ($z = 0.2248$) is 3.611 kpc/arcsecond \citep{cosmocalc}.
Note also that we adopt two key redshift values throughout this paper. One is the {BCG} redshift, z = $0.22331 \pm 0.00024$ \citep{sdss}, {and the other is} the mean A2261 cluster redshift, {$0.2248 \pm 0.0002$ (based on over 250 confirmed cluster members)}. For BCG-related calculations we use the value of 0.22331, while for cluster-wide computations, we use 0.2248. The radial velocity offset of the A2261 BCG relative to the cluster mean redshift is $-365$ km s$^{-1}$. 
{
This velocity offset is significantly larger than the errors (given above) in either the BCG redshift ($\pm 72$ km s$^{-1}$) or in the mean cluster velocity ($\pm52$ km s$^{-1}$). The BCG offset from the cluster mean velocity is, however, significantly smaller than the observed velocity dispersion of A2261, which has been measured to be $725^{+75}_{-57}$ km s$^{-1}$ \citep{rines2010}. BCGs are known to exhibit a smaller velocity dispersion about their host cluster mean velocity than that of the average cluster member \citep{lauer14}.
}
With the above cosmological parameters, the projected physical scale at the redshift of \src\ is 3.593 kpc/arcsecond.

\section{The Properties of the Stellar Knots in the Core of \src}\label{sec:knots}

\subsection{Hubble Space Telescope Observations}\label{sec:HSTdata}

\subsubsection{CLASH Imaging}

Photometry and morphology of the A2261 BCG and the 4 knots seen near its core are based on HST data obtained as part of the Cluster Lensing And Supernova survey with Hubble (CLASH) multi-cycle treasury program. The imaging observations were performed between 09-March-2011 and 21-May-2011 in 16 broadband filters from 2250 \AA\ to $1.6~\mu{\rm m}$. The center of Abell 2261 was observed for a total of 20 orbits. Complete details of the HST imaging data acquisition and reduction are given in \cite{clash}.

\subsubsection{New Spectroscopy}

\begin{figure}[!t]
\begin{center}
\includegraphics[trim=5mm 95mm 5mm 95mm,clip,width=1.0\columnwidth]{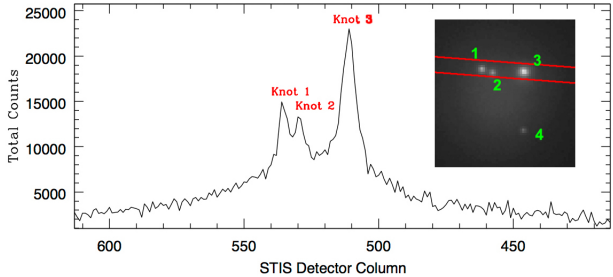}
\end{center}
\vspace{-3mm}
\caption{The summed spatial profile across 200 rows of the STIS detector demonstrating the acquisition of photons from all three targeted knots. The inset image shows the STIS $52'' \times 0.5''$ slit orientation as red lines, superposed on an HST F775W image of the core of \src. The four knots in the A2261 BCG core are numbered by their IDs from \cite{postman12}. The ORIENT angle for the STIS observations was $130.5^{\circ}$ corresponding to a position angle on the sky of  $85.5^{\circ}$. This orientation allowed spectra for knots 1, 2, and 3 to be obtained simultaneously. In this image, North is up and West is to the right.}
\label{stis_orient}
\vspace{0mm}
\end{figure}

Spectra of knots 1, 2, and 3 were obtained using the Space Telescope Imaging Spectrograph (STIS) from 31-July-2015 to 01-Aug-2015.
We used the G750L grating and the $52'' \times 0.5''$ slit. The initial pointing was centered on the brightest knot (knot 3 at R.A. $=17^h\ 22^m$\ 27\ras14 and Dec. $=+32^o\ 07'$\ 57\asec59 (J2000)).
The observations were performed with an ORIENT angle of 130\adeg5 allowing
all three knots to be observed simultaneously (Figure~\ref{stis_orient}). We integrated for a total of 8 orbits (total exposure time of 19,100 seconds) and acquired 16 exposures.
The 8 orbit integration sequence was executed as two visits, with 300 second acquisition exposures done at the start of orbits \#1 and \#5 to ensure precise slit alignment. Four small pointing shifts (each 0\asec2 in size and parallel to the long dimension of the slit) were made over the course of 4 orbits to allow for rejection of hot/bad pixels in the dispersion direction. The same shifts were repeated for the second set of 4 orbits. 

The STIS data were reduced by performing subtraction of variable herring-bone pattern noise, bias correction and pixel-based charge transfer inefficiency (CTI) correction on the raw science datasets, then running the remainder of the {\tt calstis} processing steps to produce the final CTI-corrected 2D flat-fielded images. The removal of the herring-bone pattern noise, a known issue with STIS since it resumed operations in 2001 using its Side-2 electronics, was done using the procedures outlined in \cite{jansen}. The 2D flat-fielded images were then shifted along the spatial axis using the Pyraf task {\tt sshift} to remove the offsets from the dithers done between each orbit. The individual images were co-added using the {\tt ocrreject} routine, which is designed to detect and remove cosmic rays in STIS CCD data as well as co-add input images. The benefits of using {\tt ocrreject} to combine the data are twofold: (1) an error array is produced which is proportional to the square-root of the output science image, but smaller by a factor that depends upon the square-root of the number of non-rejected input values used to compute the science pixel values and (2) it allowed us to use all 16 exposures to ensure robust CR-rejection. The final 1D spectrum for each knot was extracted using the {\tt x1d} routine which also applies the STIS flux and wavelength calibrations to produce a co-added 1D spectrum in F$_\lambda$ units of erg s$^{-1}$ cm$^{-2}$ \AA$^{-1}$. The wavelength range of the extracted 1D spectra runs from 5257 \AA\ to 10249 \AA, with a dispersion of 4.87 \AA\ per pixel.

The average signal-to-noise ratio (S/N) per resolution element (resel) in the 1000 \AA\ wide range centered on the redshifted NaD doublet feature (centered at 7218 \AA\ at the  redshift of $z=0.2248$ for A2261) for the extracted spectra from knots 1, 2, and 3 is 8.6, 7.7, and 10.7, respectively. The predicted {per-resolution-element} S/N estimates {from the HST exposure calculator} were about a factor of 1.8 higher than the achieved values. The large difference between the observed and predicted S/N is primarily due to the significant number of pixels lost to cosmic ray contamination (even with 16 independent exposures) and, to a lesser degree, by residual noise left after the CTI and pattern noise corrections are applied. However, without applying those corrections the observed S/N would have been much worse.

\subsection{Central Stellar Velocity Dispersions of the Knots}

We fitted the extracted, normalized STIS spectra to estimate the stellar velocity dispersion in the knots.  We used pPXF, the penalized pixel fitting code of \citet{2004PASP..116..138C}.  
For each knot spectrum we logarithmically rebinned in wavelength so that it matched that of the MILES library spectra \citep{2010MNRAS.404.1639V}.  To match the spectra, we convolved the higher resolution spectra with a gaussian with variance equal to the difference in variance of our observations and the template spectra.  
We tried various schemes of binning the spectrum but found they made little difference to our final results.  Thus the effective resolution is $250\,\mathrm{\kms}$.    We similarly tried a variety of multiplicative and additive polynomials to account for any residual continuum shape but there was no significant improvement to the fits.  Because of the low S/N of our data, we fit only the first two moments (velocity, $V$, and velocity dispersion, $\sigma$) of the line-of-sight velocity distribution relative to the recessional velocity of A2261.

A major concern for our fits to data of low $S/N$ is spurious effects of template mismatch.  In order to take this into consideration we used the full MILES library. By using the entire library, we are ensuring that whatever the stellar population of the knots, we have included the proper stellar template. If we had not included the correct stellar template, then at low S/N the penalized pixel fitting method is more likely to find spurious solutions.

A second concern for fits to low $S/N$ is proper accounting of measurement uncertainties in the spectrum.  We ran a series of Monte Carlo simulations to create realizations of the spectrum based on the 1D error spectrum after rebinning. The simulated data were fitted for each knot.  For each knot spectrum, we ran 1000 realizations and took the median and 68\% interval for the $V$ and $\sigma$ parameters of the fit, weighting them by the reciprocal of the square of the formal fit uncertainties.  This weighting scheme ensures that poor fits have less influence on the final results than good fits.  All had median results for the velocity dispersion to be very similar to the best-fit results.

\begin{figure}[!t]
\begin{center}
\includegraphics[width=1.0\columnwidth]{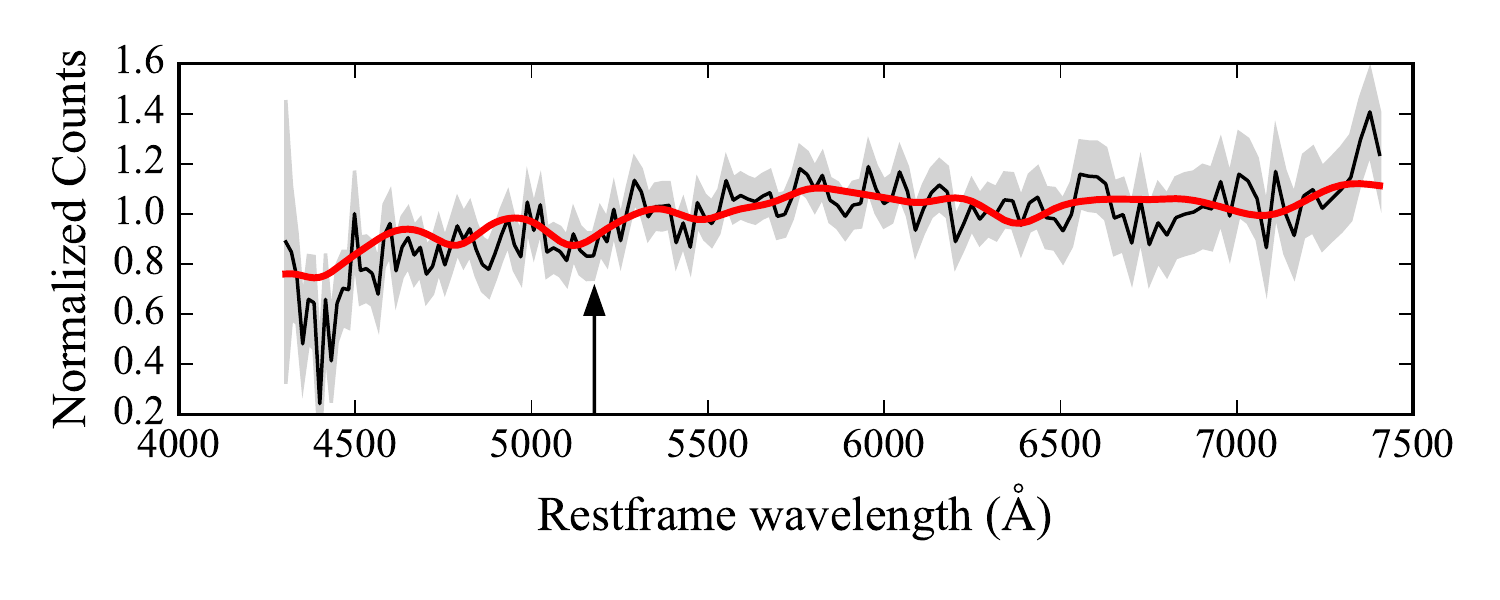}\vspace{-4mm}
\includegraphics[width=1.0\columnwidth]{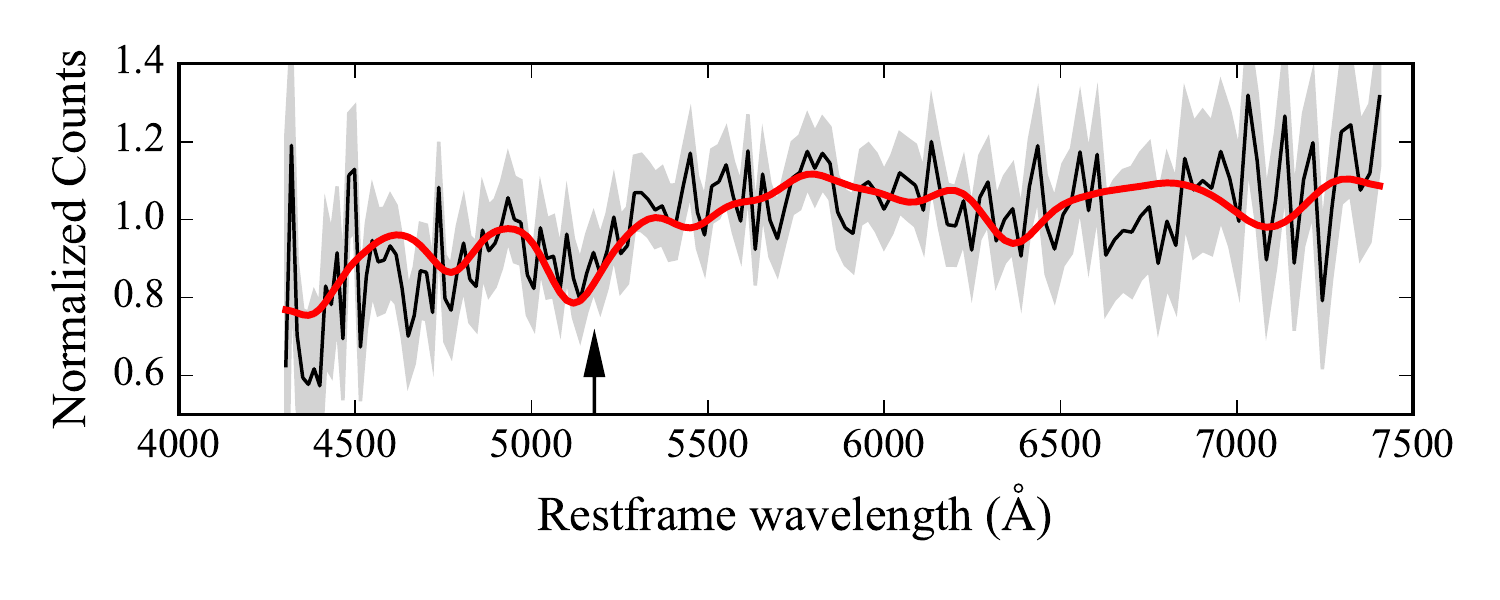}\vspace{-4mm}
\includegraphics[width=1.0\columnwidth]{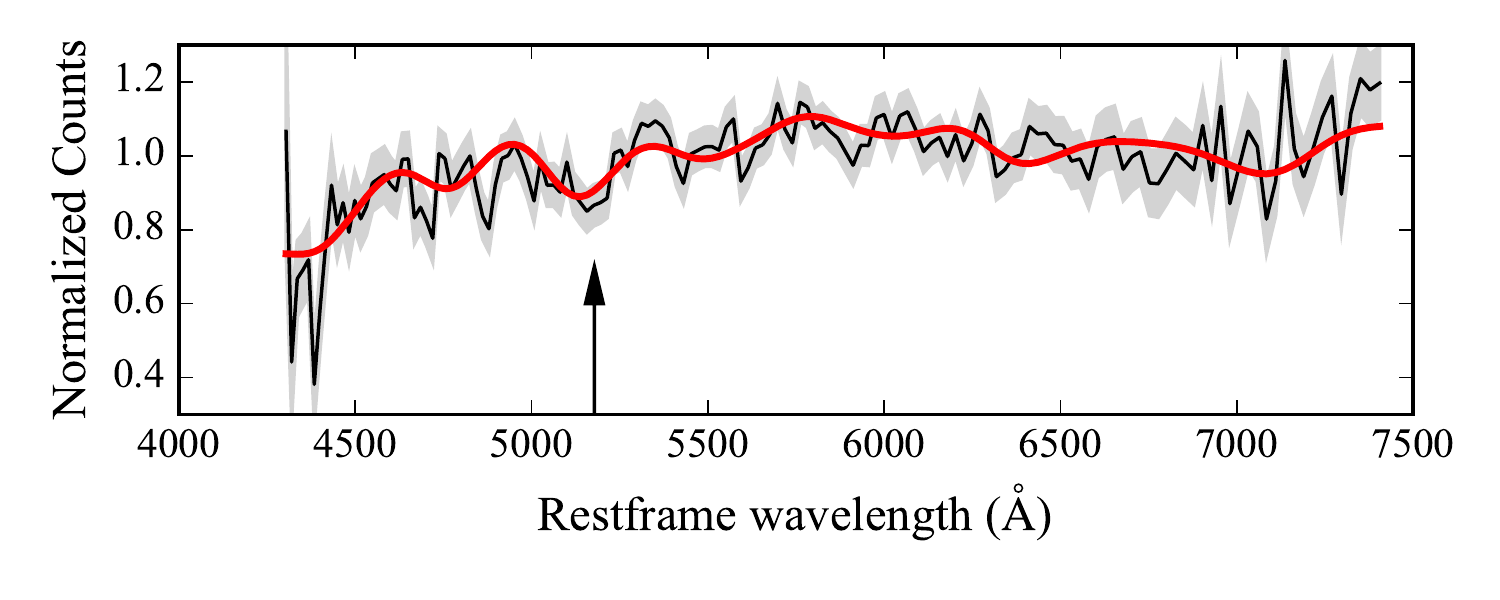}\vspace{-4mm}
\end{center}
\vspace{-2mm}
\caption{The normalized object spectra (black lines) and the 68\% confidence intervals on the observed spectrum fluxes (grey shaded areas) are shown for A2261 knots 1, 2, and 3 from top to bottom, respectively. The best fits obtained for
each knot are shown as red lines.  The arrow indicates the restframe wavelength of the Mgb spectral feature.}
\label{spectrafits}
\end{figure}

The results of our fitting are presented in Table~\ref{table:vdisp}.
Column 2 in this table is the velocity offset relative to a fiducial
redshift of $z = 0.22331$. Column 3 is the derived velocity dispersion.
For each knot, the results are expressed as the median values from our Monte Carlo simulations with 68\% interval of the parameter distributions for the uncertainties.
We plot the 1D spectra with their best fits in Figure \ref{spectrafits}.

In the spectra for knots 2 and 3, we find no significant evidence for a
velocity dispersion larger than that expected for a
{galaxy} with a black hole mass of $\lesssim 6\times10^8 M_{\odot}$. 
The knot 1 velocity dispersion measurement is highly uncertain, however,
and we cannot rule out the hypothesis that
knot 1 is hosting the ejected \src\ black hole. 

\begin{deluxetable}{lcc}
\tabletypesize{\tiny}
\tablewidth{0pt}
\tablecaption{A2261 Knot Velocity Offsets and Dispersions }
\tablehead{
\colhead{ }&
\colhead{\bf Velocity Offset$^{a}$}&
\colhead{\bf Velocity Dispersion$^{a}$}\\
\colhead{\bf Object}&
\colhead{[$\mathrm{\kms}$]}&
\colhead{[$\mathrm{\kms}$]}
}
\startdata
Knot 1 & $-\phantom{1}74_{-90}^{+65}$ & $175_{-59}^{+101}$ \\
Knot 2 & $          -117_{-97}^{+98}$ & $168_{-63}^{+66}$ \\
Knot 3 & $-\phantom{1}58_{-55}^{+62}$  & $209_{-46}^{+39}$ 
\enddata
\label{table:vdisp}
\tablenotetext{a}{{The values given are the weighted medians with uncertainties coming from the weighted 16 and 84\% of the Monte Carlo distributions.  The best fit model spectra used a mixture of stellar templates from the MILES library.  Although the modeling procedure can use any of the stars from the full library, in practice only a few templates contribute the most.  The non-negligible contributions in decreasing order of relative weights (i) to knot 1 come from G 156$-$031, HD 187216, and HD 200779; (ii) to knot 1 come from UCAC2 21686546, HD 001326B, 2MASS J15182262+0206397, HD 199799, HD 042474, and HD 183324; and (iii) to knot 3 come from 2MASS J15182262+0206397, UCAC2 21686546, HD 12632, G 171$-$010, HD 097907, BD+060648, 2MASS J05240938$-$2433300, HD 042474, and HD 023194.}}
\end{deluxetable}

% New paragraph:
{
The velocity offsets of the observed knots are significantly less than the velocity dispersions of the \src\ and the A2261 cluster (see~\S\ref{sec:intro}). The knot velocity offsets are consistent with them being cluster members and they are consistent with potentially being bound to the \src\ as well. However, in \S\ref{sec:disco}, we show that the knot velocity offsets are also consistent with the range of predicted initial kick velocities for a SMBH ejection event.
}

\subsection{Stellar Population Modeling and Stellar Mass Estimates of the Knots}\label{sec:mass}

Stellar masses for the four knots are computed using the {\tt iSEDfit} package \citep{moustakas13}. Input photometry for the iSEDfit analysis was obtained from 13 broadband images obtained from the CLASH HST Treasury program.
All CLASH filters including, and redwards of, F390W were used for our SED fitting. For knots 1, 2, and 4 we used 0\asec364 (1.31 kpc) diameter apertures and for knot 3 we used a 0\asec52 (1.87 kpc) diameter aperture. A local background subtraction was performed around each knot in each filter to account for underlying sky and BCG flux contamination. To compute the stellar masses, we generate a grid of 20,000 model spectra using a range of synthetic stellar populations \citep{BC03}, two different initial mass functions [IMF]\citep{salpeter55, chabrier03}, and the Calzetti dust extinction law \citep{calzetti2000}. We then compute the grid of model photometry obtained by convolving the model spectra with HST filter response functions to generate a posterior probability distribution for the parameter space. We adopted a star formation history consisting of an initial burst followed by an exponentially decaying star formation rate (e.g., the BC03 tau model). Table~\ref{table:knotphot} gives the best-fit stellar mass estimates derived for the knots for each of the two assumed initial mass function (IMF) models. 

Upper limits on the sizes of the knots are also given in Table~\ref{table:knotphot}. The size estimates are the half-width at half-maximum (HWHM) values of the central light distribution of each knot derived from their profiles measured in a deconvolved F814W CLASH image. The image deconvolution was performed by identifying an unsaturated star in the F814W image as our PSF reference and then applying 20 iterations of the Lucy--Richardson deconvolution technique \citep{Rich72,Lucy74}. The profile for each knot was then measured using the high-resolution {\tt PROFILE} routine \citep{Lauer85} in the {\tt xvista} image processing package. The HWHM values derived for the central light distributions are almost certainly all upper limits. HWHM values in Table~\ref{table:knotphot} are given in both arcseconds and kpc, adopting the projected scale of 3.593 kpc/arcsecond at $z = 0.22331$. All four knots have central light HWHM sizes that are equal to or less than 190 parsecs. We emphasize that these HWHM values are not the same as the effective radius. For example, knot 3 is clearly a fully resolved galaxy but the HWHM of the central light distribution reveals that its nuclear light distribution remains unresolved at {\it HST's} angular resolution.

\cite{bongra16} report the existence of a 5th knot in close proximity to knot 3. We do not model knot 5 as a separate entity. \cite{bongra16} report that knot 5 is significantly less massive any of the other knots by at least a factor of 10 and is a factor of 150 times less massive than knot 3. The STIS spectrum of knot 3 would have included all the light from knot 5 as well. The S/N of knot 5 would have been far too low for its kinematics to contribute in any way to the kinematic parameters we derived for knot 3.

\begin{deluxetable*}{lccccc}
\tabletypesize{\small}
\tablewidth{0pt}
\tablecaption{Best Fit Stellar Mass Estimates and Knot Central Size Estimates}
\tablehead{
\colhead{}&
\colhead{Mass in $R_m$}&
\colhead{Mass in $R_m$}& 
\colhead{}&
\colhead{}&
\colhead{}\\
\colhead{}& 
\colhead{(Salpeter IMF)} & 
\colhead{(Chabrier IMF)} & 
\colhead{$R_m$}&
\colhead{HWHM$^a$}&
\colhead{HWHM$^b$} \\
\colhead{{\bf Object}} &
\colhead{{[$10^{10}\,\msun$]}} & 
\colhead{{[$10^{10}\,\msun$]}} & 
\colhead{[kpc]}&
\colhead{[arcsec]}&
\colhead{[kpc]}\\
}
\startdata
Knot 1 & $0.64^{+0.09}_{-0.08}$ & $0.37^{+0.05}_{-0.05}$ & 0.66 & $\le$ 0.047 & $\le$ 0.17\\
Knot 2 & $0.54^{+0.08}_{-0.07}$ & $0.31^{+0.05}_{-0.04}$ & 0.66 & $\le$ 0.043 & $\le$ 0.15\\
Knot 3 & $1.70^{+0.24}_{-0.20}$ & $0.98^{+0.15}_{-0.12}$ & 0.94 & $\le$ 0.053 & $\le$ 0.19\\ 
Knot 4 & $0.16^{+0.02}_{-0.02}$ & $0.09^{+0.01}_{-0.01}$ & 0.66 & $\le$ 0.041 & $\le$ 0.15\\
\enddata
\label{table:knotphot}
\tablenotetext{a}{The half-width at half-maximum (HWHM) of the central light distribution is derived from the deconvolved F814W CLASH image. See \S\ref{sec:mass} for details.}
\tablenotetext{b}{Assumes 3.593 kpc/arcsec at $z = 0.22331$, {i.e.\ assumes the knots reside in the BCG core.}}
\end{deluxetable*}

\subsection{Broadband Colors of the Knots}

\begin{figure}[!t]
\begin{center}
\includegraphics[width=1.0\columnwidth]{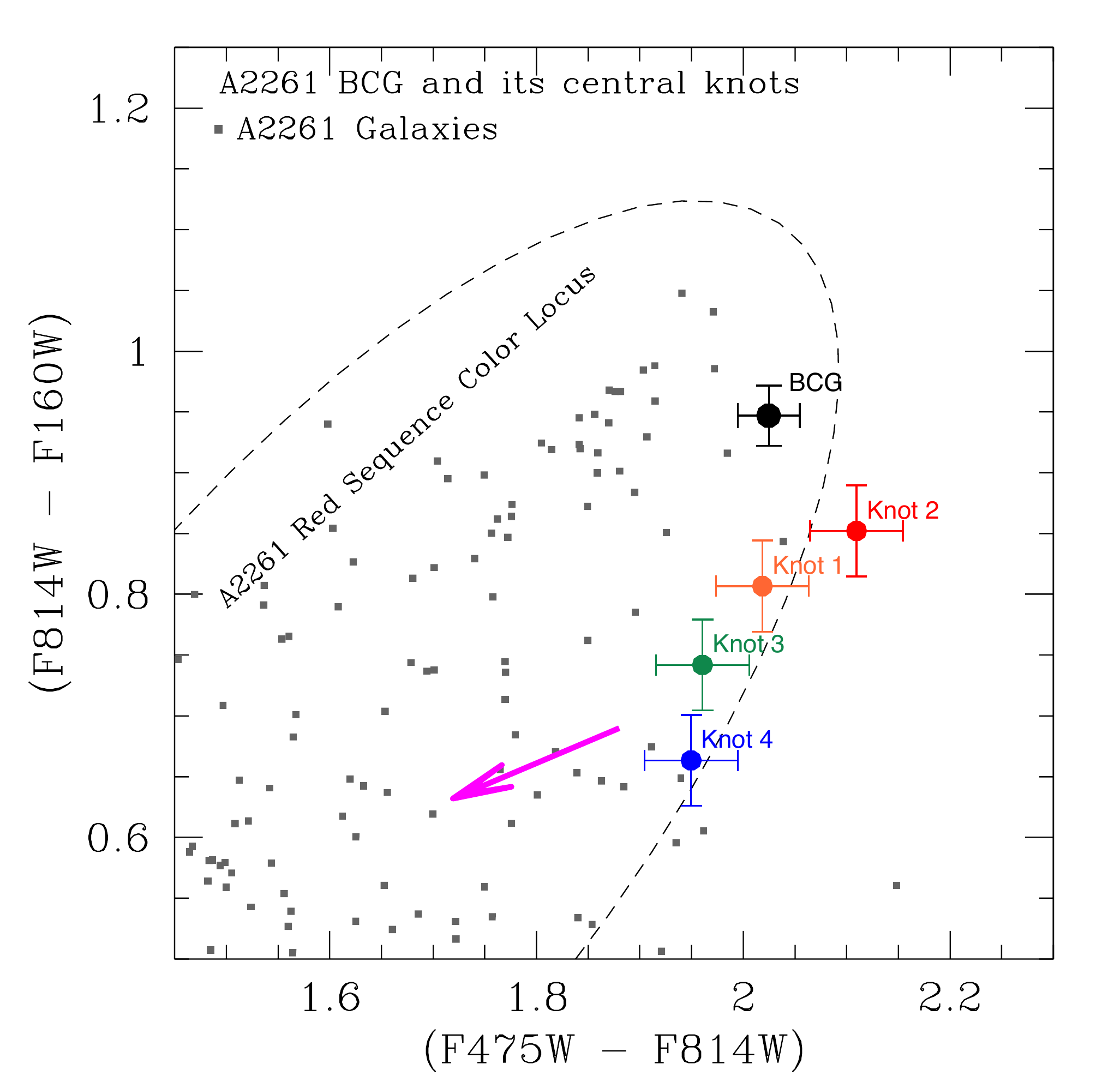}
\end{center}
\vspace{-4mm}
\caption{The color-color plot of $(F475W - F814)$ vs. $(F814W - F160W)$ for A2261 knots 1, 2,  3, and 4 as well as that for the \src. The error bars are 1-$\sigma$ values. Colors for other galaxies in the CLASH observations of A2261 (Conner et al. 2017; private comm.) are shown as grey points. The ellipse identifies the outer boundaries of the color-color parameter space of the A2261 red sequence. The arrow indicates the direction of a typical elliptical galaxy color gradient going from the knot radius of $\sim0.25$ kpc to a scale length of 2.5 kpc. {The colors of the points for knots 1--4 are simply to guide the eye.}}
\label{knotcolors}
\vspace{3mm}
\end{figure}

We measured the observed $F475W - F814W$ and $F814W - F160W$ colors for knots 1 through 4 while preparing the input photometry for our stellar mass estimation described in \S\ref{sec:mass}. The colors of the four knots and that of the BCG (within its central 23.5 kpc) are shown in Figure~\ref{knotcolors}. The colors shown are corrected for local Galactic extinction using the results from \cite{schlegel98}. Knot 4 is the bluest and Knot 2 the reddest. 
The four knots are located near the periphery of the red sequence in this color--color space. All four knots are at the red end of the $(F475W - F814W)$ red sequence color distribution although they have $(F814W - F160W)$ colors that are consistent with the bulk of the red sequence galaxies. While we cannot put strong constraints on the age or metallicity of the knots, the SED fitting results suggest that knots 3 and 4 are likely to have subsolar metallicity. Specifically, the probability that the stellar populations of knots 3 and 4 are subsolar are 0.93 and 0.91, respectively, for the Salpeter and Chabrier IMF. Our models do not provide any significant constraints on the metallicity of knots 1 and 2.

If the knots are the stripped nuclei of initially larger early-type cluster galaxies then they would be expected to be somewhat redder than the average red sequence cluster member. Elliptical galaxies are observed to have color gradients with the general trend of the stellar population becoming slightly bluer as one moves out to larger radii. Specifically, observed color gradients in ellipticals \citep{tamura2000} are $\Delta(F606W-F814W)/\Delta {\rm log}(r) \sim -0.06$ in the redshift range $0.2 \lesssim z \lesssim 0.3$. The origin of color gradients in elliptical galaxies appears to be predominantly due to a decrease in the metallicity of the stellar population with increasing radius from the galaxy center. The arrow in Figure~\ref{knotcolors} shows the direction in which elliptical (F475W-F814W) and (F814W-F160W) colors would change, on average, as one moves from a region corresponding to the size scale of the knots ($\sim 0.25$ kpc) out to a radius of $\sim 2 - 3$ kpc, corresponding to the effective radius of a cluster elliptical. The trend denoted by the arrow corresponds to $\Delta (F475W-F814W) = -0.16$ mag and  $\Delta(F814W-F160W) = -0.058$ mag for $\Delta (F606W-F814W)/\Delta {\rm log}(r) = -0.06$ and $\Delta {\rm log}(r) = 1$. If the knots are the nuclei of stripped early type galaxies then their progenitors are consistent with being red sequence members of Abell 2261.

\subsection{Knots 2 and 3 are likely %dwarf 
{small} cluster galaxies; Knots 1 and 4 remain inconclusive.}\label{sec:knotcolors}

\begin{figure}[!t]
\begin{center}
\vspace{0mm}
\includegraphics[trim=3mm 6mm 10mm 5mm,clip,width=1.0\columnwidth]{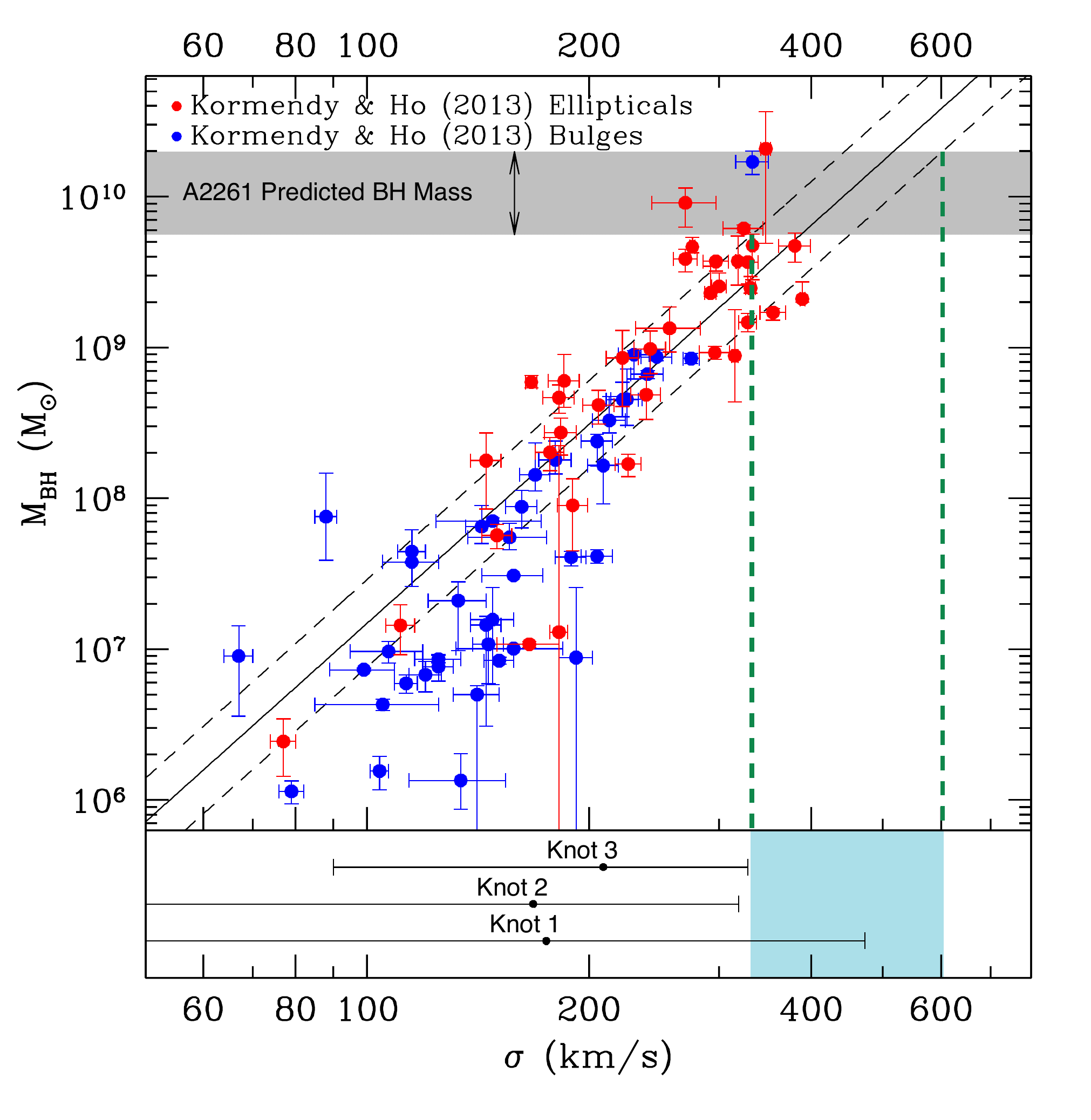}
\end{center}
\vspace{-4mm}
\caption{Top: The black hole mass,  M$_{\bullet}$, as a function of central stellar velocity dispersion, $\sigma$, is shown for the \citet{korho13} sample. Their best-fit relation and its $1-$sigma scatter are shown as black lines. The red and blue points are the measurements for elliptical galaxies and galactic bulges, respectively. The mass range for the nuclear black hole in \src\ is shown as the horizontal gray band. The implied limits on the velocity dispersion from this range of M$_{\bullet}$ values are projected down onto the x-axis by the green dashed lines. Bottom: The observed knot stellar velocity dispersion measurements from this work are shown along with their 95\% confidence limits. The blue shaded region shows the expected range in $\sigma$ for the estimated range in mass for the \src\ black hole.}
\label{bhmass-sigma}
\vspace{3mm}
\end{figure}

As Tables~\ref{table:vdisp}~and~\ref{table:knotphot} suggest, knots 2 and 3 are photometrically and kinematically consistent with the properties of 
%dwarf 
 {small} galaxies. Their velocity dispersions are between {168 and 209} km s$^{-1}$ and they have stellar masses of $3.1 \times 10^9$ and $9.8 \times 10^{9}$ M$_{\odot}$, respectively (using a Chabrier IMF). The velocity dispersions of knots 2 and 3 are inconsistent at the $\gtrsim2\sigma$ level with systems containing a black hole with mass comparable to that predicted for \src, \mbox{$M_{\bullet} \gtrsim 5.6 \times 10^9$\,M$_{\odot}$}, which would have yielded an observed stellar velocity dispersion in excess of $330$\,km\,s$^{-1}$ \citep{korho13}. Figure~\ref{bhmass-sigma} shows the predicted range in  velocity dispersion for the estimated range in the \src\ black hole mass discussed in section~\ref{sec:intro}.
As noted above,
the large errors in the velocity dispersion of knot 1 mean that we cannot
rule out the hypothesis that it is hosting a high-mass BH.

\section{The Relic Radio Source in \src}\label{sec:radio}

\subsection{Very Large Array Observations}\label{sec:VLAdata}

As reported by \citet{postman12}, \src\ was detected by the FIRST survey \citep{FIRST} as an unresolved 1.4\,GHz radio source with a flux of 3.4\,mJy, 1\asec6 west of the galaxy's position; this is just under a factor of two of the galaxy core radius. 
We thus collected wide-band continuum data from the VLA to obtain:
\begin{itemize}
\item Accurate reference frame ties to the available \mbox{Subaru} optical image \citep{postman12} to determine whether the offset position was genuine;
\item A resolved image of the radio feature to assess its nature morphologically; and
\item An estimate of the jet age through radio spectral aging \citep[\eg][]{carilli+91}.
\end{itemize}

\begin{table*}
\centering
\scriptsize
\caption{VLA Imaging Results}% {Beam/source P.A.'s and numbers need to be triple-checked.}}
\begin{tabular}{cccccccccc}
\hline
        & $f_{\rm center}$ &  {\bf $\sigma$} & {\bf Peak}        & {\bf Int.} & {\bf Beam size} & {\bf Beam} &{\bf Source Size} & {\bf Source} & {\bf Linear}\\
{\bf Band} & {\bf (GHz)} & {\bf ($\mu$Jy)}& {\bf ($\mu$Jy)} & {\bf ($\mu$Jy)} & {\bf (mas)}& {\bf P.A. (deg)} &{\bf (mas)} & {\bf P.A. (deg)} & {\bf extent (kpc)}\\
\hline
% These are flux/beam size numbers from **POOR** fits.
C & 5.873 & 2.1 & $132.4\pm6.8$ & $483\pm31$ &$530\times360$& 68.0 & 950$\pm$$70\times470$$\pm$60& $135\pm5$&3.45\\
X & 9.937 & 1.2 & $23.4\pm1.9$   & $158\pm14$ & $340\times220$ & 72.0& 840$\pm$$80\times490$$\pm$60& $126\pm9$&3.05\\
% Where did I get the below numbers? Is it from kvis or something?!?
%C & 5.873 & 2.1 & 142.1 & 10.1 &$530\times360$& 68.0 & 950$\pm$$70\times470$$\pm$60& $135\pm5$\\
%X & 9.937 & 1.2 & 27.3   &  2.63 &  & xxx& xxx&  xxx\\
\hline
\end{tabular}\label{table:radio}
\vspace{1mm}
\end{table*}

Data were collected under project code VLA/15A-061 in A-configuration in two frequency ranges. The first was 3.976--8.024\,GHz (C-band, $f_{\rm avg}=5.873$\,GHz after flagging, hereafter ``6\,GHz''), which provided 0\asec3 imaging resolution. The second was 7.976--11.896\,GHz (X-band, $f_{\rm avg}=9.937$\,GHz, hereafter ``10\,GHz''), which provided 0\asec2 resolution. 
%These frequencies both have resolutions well-matched to the available Hubble Space Telescope images. 
Both frequency set-ups used standard 2\,MHz channel widths and 2\,s sampling. Standard primary calibrator 3C286 was used for flux density and bandpass calibration, and source J1735+3616 was used as a phase calibrator. To gain sufficient sensitivity, observations were performed in two epochs and then concatenated to create a single data set per frequency band. Given the extended emission in the source, we do not expect any significant intrinsic variability between the two epochs.

The 6\,GHz observations were performed on 11 July 2015, and 13 July 2015, giving a total of 6.27\,h on-source time.
At 10\,GHz, the observations were made on 25 June 2015 and 27 June 2015; 6.70\,hours were spent on-source at this frequency. We utilized the standard VLA pipelines to calibrate the data, and performed manual flagging and imaging with the {\tt CASA} software package. Table\,\ref{table:radio} provides the resulting image noise, measurement, and beam parameters.

%We imaged he VLA observations 
%^&We employed a full bandwidth of 2.048\,GHz, broken into 16 spectral windows, each containing 64 channels of 2\,MHz width. The centre frequency is 8.936\,GHz. The full 27-antenna array was online for our observations. The observations of \src\ were centred at right ascension $\alpha=$~17:22:27.18, declination $\delta=$~+32:07:57.30; this was the position noted by \citep{postman12} to be the galaxy centre.
%The observations and data reduction followed standard VLA phase-reference experiment procedures. 3C286 was used as a primary flux and bandpass calibrator, and standard VLA calibrator J1735+3616 was observed every 10\,min for $\sim$50\,s to provide phase-reference calibration. The total time on \src\ was !!!N!!! minutes. Standard calibration and imaging were performed using the {\sc casa} software package.

%Image properties are given in Table\,\ref{table:radio}. In Figure \ref{fig:contopt}, we show the 
%, and several views of the resulting images are shown in 
%Figs.\,\ref{fig:contopt}--\ref{fig:last}.

\begin{figure}
\centering
%\vspace{3mm}
\includegraphics[angle=270,width=0.95\columnwidth]{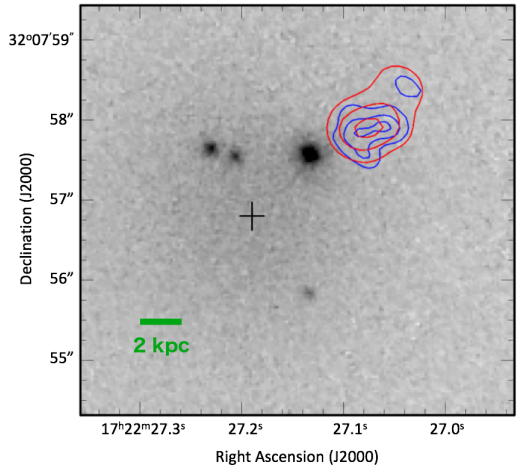}
\caption{The 6\,GHz (red) and 10\,GHz (blue) contours overlaid on the HST F814W image of the optical knots. Radio contours are at 30, 60, and 90\% of the respective image peaks. The photometric centroid of the \emph{core} model of \citet{postman12} is marked with a cross.}\label{fig:contopt}
\vspace{1mm}
\end{figure}

\begin{table*}
\centering
{\scriptsize
\caption{Astrometric Verification. Parenthesis indicate error on last digit.}
\begin{tabular}{llllll}
\hline
&\multicolumn{2}{c}{\bf Radio (J2000)}  &  \multicolumn{2}{c}{\bf Optical (J2000)}\\
{\bf Object} & {\bf R.A.}  & {\bf Dec} &  {\bf R.A.} &  {\bf Dec} & {\bf Offset ($''$)}\\
\hline
%Bright, resolved source to northwest (position of its non-extended/core component):
{Field} Spiral & {17:22:17.01(2)} & {$+$32:09:12.8(5)} & 17:22:17.016(7)  & $+$32:09:13.1(1) & {$0.3\pm0.5$} \\
Elliptical galaxy & 17:22:26.920(3) & $+$32:06:36.793(3) & 17:22:26.919(7) & $+$32:06:36.8(1) &  $0.01\pm0.15$ \\
%\hline
\src\ radio (10\,GHz)/optical core &17:22:27.072(3) & $+$32:07:57.8(2) & 17:22:27.190(7) & $+$32:07:56.8(1) & $1.80\pm0.25$\\
\hline
\end{tabular}\label{table:astrometry}
\vspace{3mm}}
\end{table*}

\subsection{The radio source is offset from, but related to, \src's core.}
FIRST images {indicated, to low significance, that there was a radio component offset from the BCG's core.} Our new higher-resolution images (Figure \ref{fig:contopt}) confirm this offset. We find the radio centroid to be 1\asec80$\pm$0\asec25 from the photometric center of the optical core.

To verify that an inaccurate radio/optical reference frame tie could not be the cause of the offset, in our data we identified two field sources detected in both our VLA image and the Subaru image (an R-band image from Suprime-cam; \citealt{postman12}). The positions of the optical centroid and the unresolved radio core component of these galaxies, and their radio/optical positional differences, are reported in Table \ref{table:astrometry}. In the same table we show \src's fitted radio position (fitting the X-band data for a single gaussian source), and the core's peak optical pixel location after the knots have been removed. We thus find the offset of the radio source to be significant and not due to a reference frame error.

We can also assess the probability that this radio source is a projected association and not a genuine one. The galaxy lies in the center of a dense cluster, and so general radio counts will likely underestimate this number. We thus measure the radio source count directly from the \src\ field using an image out to the full width half maximum image at C-band, {which has the larger field of view of the two frequencies.  The C-band field of view is 0.01 deg$^2$.}
%which covers a FOV of 
% 4096 pixels at (0.101/3600) deg per pixel --> radius of circle = ((0.101/3600)*4096)/2 = 0.057 deg
% pi*0.057^2 = 0.0104 deg^2 field of view.
% 3 / 0.0104 = 288
% Poisson error on 3: (0.003 - 6.01) (0.0027 - 6.008)
% (0.27 - 601 per sq deg)
%0.01\,deg$^2$ (wider than that of X-band). 
{There were three objects in this field with flux densities equal to or larger than the flux density of the A2261-BCG radio source, including the A2261-BCG radio source itself.}
%There were three objects in this field, including the offset central source under consideration, at or equal to the flux of the object in question within 1$\sigma$ flux errors.
Thus, the count is $R(\gtrsim140\mu{\rm Jy})\simeq300\,{\rm deg}^{-2}$ in this field. The radio offset is 6.5\,kpc (1\asec80), which subtends a circular aperture around the core center of area $3.9\times10^{-7}\,{\rm deg}^2$. A radio source of this flux density had a probability of $1.2\times10^{-4}$ to be in the galaxy core by chance; 
% This is very close to 4.0 sigma.
we are thus confident that the radio detection is related to the core.

{One further possibility is that we are seeing a weakly lensed (magnified) background galaxy. While the source is also not radially elongated, as we might naively expect for a weakly lensed object, this might be due to structural complexity in the lensed galaxy and/or the BCG itself. However, the radio source lies within a nominal Einstein radius for this object ($\sim$8$''$ assuming a point mass, $z=0.3$ of the background galaxy, and $M_{2500}\simeq10^{14}\msun$ for the BCG), making significant magnification unlikely. Future work will have to investigate this possibility further using more complex modeling of the mass distribution in the large, flat core in which this radio source resides.}

\subsection{The radio source is a relic with no presently active core.}\label{sec:radiorelic}
%We must take care to not over-interpret the faint radio component associated with \src, as its precise progenitor is ambiguous. However, as discussed here, it is clearly a relic (no longer active) region of radio emission, and likely represents a previous bout of AGN activity in this galaxy's core. 

\begin{figure}
\centering
\includegraphics[angle=270,trim=3cm 2.8cm 3cm 3cm,clip,width=1.00\columnwidth]{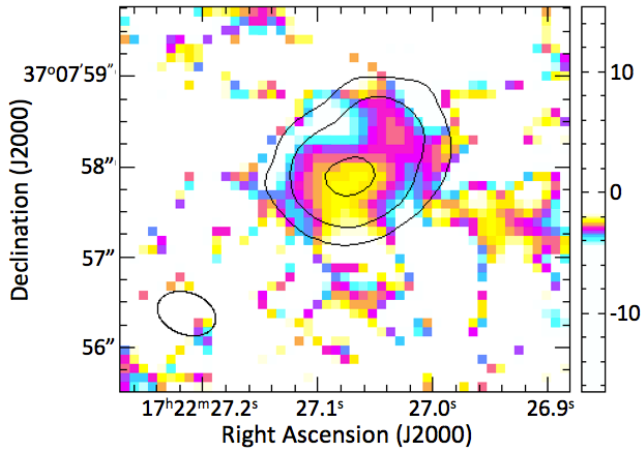}
\caption{
A two-point spectral index map of the field based on our 6 and 10\,GHz data, where spectral index $\alpha$ is defined as $S\propto f^{\alpha}$. Contours are shown for the tapered 6\,GHz data to give a reference for source placement. The lowest contour is at three times the root-mean-square noise in the 6\,GHz image; toward the edges of the object the contours become non-physical. The bulk of the source has a spectral index of around $-2$, while its spectral index steepens somewhat towards the north-western spur. These indices are indicative of aging synchrotron emission, implying the source is not currently active. No core emission was detected to the 3$\sigma$ flux limit of 6.3\,$\mu$Jy and 3.6$\mu$Jy in the C- and X-bands, respectively.}\label{fig:spix}
\label{fig:last}
\vspace{2mm}
\end{figure}

{The radio luminosity of this source is $\sim5\times10^{23}$\,W\,Hz$^{-1}$, as determined from the FIRST flux density.}
%The 1.4\,GHz FIRST flux gives a luminosity for this source of .
% In cgs units this luminosity is 7x10^39 erg/s.
%While the integrated brightness temperature is quite faint, 
The most compact component at C-band has a brightness temperature of $T_{\rm B} > 1.4\times10^3$\,K. 
%indicate that this radio feature is AGN-related.
The emission appears to cover a projected linear extent of around $1 \times 3.5$\,kpc, with a projected centroid offset of 6.5\,kpc from the core.
%and $T_{\rm B10} = !!!N$\,K at 6 and 10\,GHz, respectively. These values are sufficiently high to note that the source must be an 
Using our new flux measurements and previous lower-frequency measurements by \citet{hoganthesis}, we find a spectral index of $\alpha=-1.5\pm0.1$, in agreement with their fit of the lower-frequency data only.
Figure \ref{fig:spix} shows a resolved two-point spectral index map of our 6 and 10\,GHz data across the full extent of the radio emission. Although there is substantial error on the indices in this map ($\sim$0.5--1 on-source), it is clear that there are no flat-spectrum sub-components in the observed emission that could feasibly represent a presently active nuclear black hole.

\begin{figure}
\centering
\includegraphics[trim=20mm 20mm 20mm 20mm,clip,width=0.95\columnwidth]{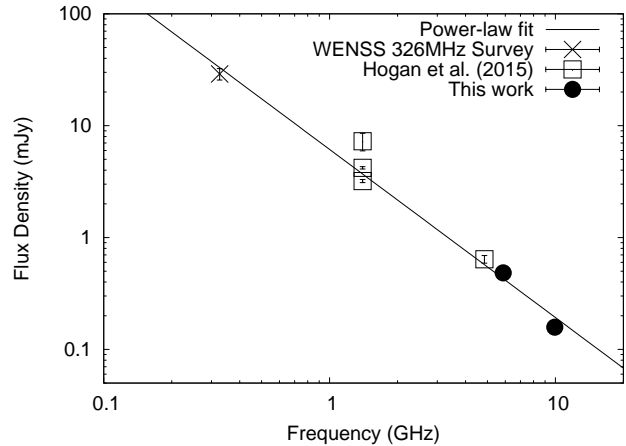}
\caption{The spectral energy distribution of the \src\ radio source. Our measurements indicate a consistency with a single power-law with a steeply declining spectrum. The steep spectral index of this object and the lack of spectrally (or spatially) resolved core component implies that the radio source is in fact no longer active. {Note that our WENSS survey data point differs from that of Hogan et al.\ because for this data point they used the radio source at the center of the Abell 2261 cluster, rather than the BCG itself.}}\label{fig:hoganspectrum}
\label{fig:last}
\vspace{2mm}
\end{figure}

%Figure \ref{fig:hoganspectrum} shows the fitted low-frequency spectrum of \src\ from \citet{hogan+15} (hereafter H15), with our new higher-frequency data points points included. The H15 study aimed to perform core/non-core component modelling for AGN in the nuclei of brightest cluster galaxies, and could only put a limit on the core emission from this source, assuming it is an AGN. Our data demonstrate that a single power-law model continues to fit this source across nearly two decades in frequency. Our 10\,GHz data point furthermore pushes a core component limit downward by a factor of $\sim$4, implying that any feasible core would contributes well under 1\,$\mu$Jy of flux density at GHz frequencies. 
%{I'd like to make a statement here about how this would be unexpected if such a whoppingly huge black hole in this galaxy were at all active. Need to figure out how to make that statement more quantitative, however.}
The spectrum and morphology of this emission are strong evidence that the radio object is in fact relic emission, whose steep slope is indicative of an aged population of synchrotron-emitting electrons \citep[\eg][]{Scheuer+Williams1968,carilli+91}. Relics encompass a range of morphologies and properties, however as argued by \citet{feretti+12}, compact relics without sharp edges that are resident in the cores of galaxies are interpreted either as a fading jet component from a source that is no longer active, or as a ``radio phoenix,'' which can arise due to compression shocks during a merger that re-energizes plasma from a previously emitting radio jet \citep[\eg][]{ensslin+98}.  However, the central location of this emission, and the lack of other larger-scale emission directly related to the BCG, appear to imply that this relic marks an active nucleus shut-off event. We cannot isolate what caused the central object to cease its activity; it could simply have run out of fuel, or it could be a merger- or recoil-induced quenching of the radio emission.

The limited size of the emission is suggestive that either the black hole's activity was short-lived (such that it had insufficient time to grow to a large radio source), and/or that any radio emission extended beyond the central regions of the galaxy faded more quickly due to lower pressure and magnetic field strength (and thus now only the central regions are visible).

A last, but more tenuous, possibility for the origin of the relic in this core is that it is old emission that originated from the extended active galactic nucleus that exists approximately 2\amin5 northwest of the BCG, in a direction of about 45$^\circ$ from north. {Note that this is the object referred to as a ``field spiral'' in Table~\ref{table:astrometry}, as identified by the Subaru image.} This plasma may have been re-energized due to sloshing or shocks from ongoing cannibalization in the \src\ core.  The other radio object, at J2000 RA, Dec J17:22:17, +32:09:13, has emission that extends from its central bright source outwards in a general direction towards the core of \src, as shown in {the maps of} \citet{sommer+17}. The direction of the core of that object in general agrees with the offset direction of the radio source with the \src\ photometric centroid.
If one draws a line from the BCG center through its resident relic, it points generally down the barrel of the other galaxy's axis and through its central bright source. While this is suggestive, it does not imply direct causality.

We also explored whether this object could be explained as a medium-sized symmetric object (MSO), a kiloparsec-scale radio source representing an intermediate stage in the evolution of extragalactic radio sources from parsec scales (compact symmetric objects [CSOs]) to the classical Faranoff-Riley~II sources \citep{1995A&A...302..317F,1996ApJ...460..634R}. 
This interpretation would favor the presence of an offset SMBH, resident somewhere within the offset location of the radio emission. The principal objection to this explanation is that this source is significantly under-luminous relative to the expectations for MSOs.  CSOs and MSOs show well-defined scalings of luminosity and size, with objects having a size scale of order 1~kpc expected to have spectral luminosities of order $10^{27}$~W~Hz${}^{-1}$ at~5~GHz \citep{1995A&A...302..317F, 1996ApJ...460..634R, 2016MNRAS.459..820T}. By contrast, the spectral luminosity of this source at 5~GHz is approximately $7 \times 10^{22}$~W~Hz${}^{-1}$.  While the scatter in this relation is large ($\sim 1$~dex), it is insufficient to accommodate this difference. A secondary objection is that this source is spectrally anomalous relative to most CSOs and MSOs, which tend both to have a flat-spectrum core and only a small fraction of the source with a spectral slope as steep as $-1.5$.  Even in the regions where spectral indices are steeper than this, these regions are typically at the edges of the source, rather than being 
the entire source \citep{2016MNRAS.459..820T}.
Finally, CSOs and MSOs tend to be made up of a set of compact objects rather than dominated by a continuous, resolved component, as this object is.
%$>$11.5 kyr old if you say material is travelling outward at speed of light from its center. However, this material is resolved, so seems diffuse! CSOs tend to be made up of compact objecst rather than dominated by resolved emission like this (e.g. this thing is also 1kpc wide)

\subsection{How old is the radio relic?}\label{sec:relicage}
We can place a limit on the age of the radio component in this galaxy using synchrotron aging models, \eg\ \citet{harwood-spectral-aging}. These typically model the broad-band radio spectrum with a double power-law, with a break frequency ($f_{\rm b}$) above which the aging electrons have a steep spectrum, and below which the spectrum exhibits a more standard ($\alpha\simeq-0.7$) injection value. In these models, $f_{\rm b}$ and the ambient magnetic field in the radio lobe ($B$) can be used to estimate the time since last electron acceleration: $t = 1610\,B^{-3/2} f_{\rm b}^{-1/2}$\,Myr, with $B$ in $\mu$G, and $f_{\rm b}$ in GHz \citep[c.f.][]{Pacholczyk70,ensslin+98}.

\begin{figure}
\centering
\includegraphics[trim=1.9cm 2.1cm 1.4cm 2.6cm,clip,width=0.97\columnwidth]{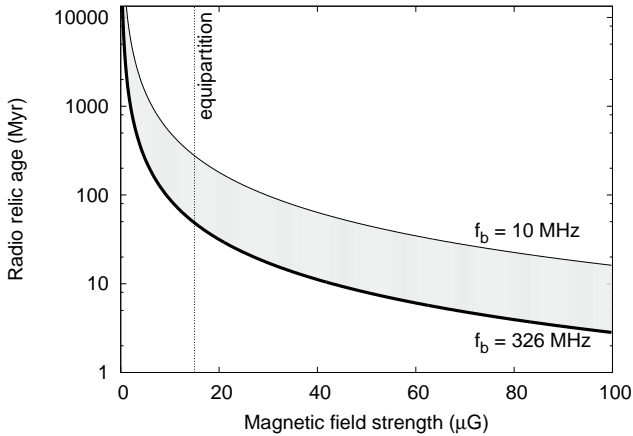}
\caption{{Radio relic age as a function of local magnetic field strength. The 326\,MHz line represents the largest plausible spectral break frequency based on the observed spectrum. 
%For an equipartition condition, the inferred magnetic field at this source is shown (\S\ref{sec:relicage}). This sets a nominal limit 
Equipartition sets a lower limit on the source age to $>48$\,Myr, with an estimated range $\sim$50--200\,Myr.}}
%For a conservative limit on cooling core cluster central galaxy magnetic fields of $\lesssim$50\,$\mu$G, we can limit the age to $>8$\,Myr.}
\label{fig:relicage}
\vspace{0.1mm}
\end{figure}

From {Figure \ref{fig:hoganspectrum},} there appears to be no clear low-frequency spectral break down to the lowest data point at 326\,MHz \citep{wenss}. Given the steep spectrum across these frequencies that is well-fit to a single power law, we therefore infer that the break frequency is $f_{\rm b}<326$\,MHz. We also require a magnetic field to translate this to a limit on relic age. {In the absence of any determinations of the magnetic field in \src, Figure~\ref{fig:relicage} shows} the inferred range of relic age estimates for break frequencies ranging from 10\,MHz to 326\,MHz, as a function of magnetic field strength. The implication is that the relic must be very old ($>$10's of Myr), have an uncharacteristically high magnetic field, or both. The very low spectral break implied for this source implies a magnetic field in the relic on the order of $\gtrsim$2\,$\mu$G if the relic is aged 1\,Gyr, and $\gtrsim$15\,$\mu$G for a more typical relic age of 50\,Myr.
%These field values are roughly consistent with other analyses of magnetic fields associated with radio sources in cooling-core cluster BCGs, if we assume the magnetic filling factor is much less than unity \citep[\eg][]{eilek-owen2002,govoniferetti04}.
 %The implied strength of the magnetic field supplies further support that in fact the radio relic is resident in the galaxy core, where the field might reach tens of $\mu$G, rather than further out in the cluster and simply in projection with the core. %!!! This is not necessarily true, as magnetic field might simply be stronger local to the radio source instead of generally in the galaxy core.

{
We can estimate the magnetic field by assuming equipartition between magnetic energy and relativistic particle energy. Following \citet{Pacholczyk70}, and assuming that there is equal energy in heavy particles and a filling factor of 1 (i.\,e.\ the magnetic field occupies the whole region rather than in filaments), the minimum value of the total energy is reached when:
\begin{equation}
B_{\rm min} = (9\,c_{\rm 12})^{2/7}R^{-6/7}L^{2/7}
\end{equation}
where $R$ is the source depth, $L$ is the luminosity, and where in CGS units $c_{\rm 12}=1.7\times 10^8$ is a factor that depends on the radio spectrum and a lower and upper frequency cutoff, assumed here to be 10\,MHz and 100\,GHz, respectively \citep[\eg][]{harris93}. Note that 10\,MHz is a standard choice for a lower cutoff, while the choice of upper frequency cutoff does not greatly effect the results (between choices of 1\,GHz and 100\,GHz, $B_{\rm min}$ varies by $<$$0.5\,\mu$G). For radio relic depth of $\sim$3kpc, we find the equipartition magnetic field is $\sim$15$\,\mu$G. Note that because the equipartition field is most sensitive to the size of the source, and because this radio source is compact when compared to other radio relics, the magnetic field in this object is about one to two orders of magnitude higher than more diffuse radio relics \citep[\eg][]{giovannini93}.  This is large, but perhaps unsurprising given the location of this relic in proximity to the core of this BCG.}
%Based on the discussions of \citet{govoniferetti04}, as a conservative limit we can acquire a very conservative lower limit on the radio relic's age by limiting the magnetic field strength to $B<50\,\mu$G, placing a rough limit on the radio relic age to be $\gtrsim$8\,Myr. }
%This is the conservative limit we will use in the discussion throughout the rest of this paper.
 %For the remainder of this paper, we assume a magnetic field strength
 %If relic is age of universe, the mag field has to be 0.4 uG. Discussion of magnetic fields in different media. Likely age is a few 10's of Myr.

%, and several views of the resulting images are shown in 
%Figs.\,\ref{fig:contopt}--\ref{fig:last}.

\subsection{The jet relic does not mark an optical knot.}
It is tempting to trace the jet axis's alignment back to \src's optical core position or to one of the knots. However, based on the likely age of the jet and mobility of the knots, we conclude that the relative positioning of the radio and optical components cannot be meaningfully interpreted.

%{leave it at that or is it worth getting into some ``sloshing'' discussion here as we have previously discussed?}
 %MENTION SLOSHING... it happens on the scale of a few crossing times which can be estimated and that number should be there.
 
%\noindent {\sc Main points to make here:}
%\begin{itemize}
%\item Discuss coincident field source positions (including a table) to demonstrate that {\bf the offset of the radio source is real}.
%\item {\bf Probabilistically, the radio source is associated with the BCG.} 
%\item Present Fig.\,\ref{fig:contopt} and a spectrum.
%\item Discuss brightnesses and compactness and make argument that {\bf this is certainly an AGN and not something else}
%\item The source is steep spectrum, and thus {\bf the jet is no longer active}.
%\item {\bf This is not a CSO.}
%\item Discuss morphology and alignment to the knots; does the jet originate in one of the knots? Basic answer we will conclude is: {\bf the morphology is sufficiently complex that we can't determine where the jet originated from. But, the evidence points to the fact that the jet is no longer active and thus was disrupted something like $10^6$ to $10^7$ years ago, otherwise we shouldn't be seeing this source at all, and with such a steep spectrum. Thus, the morphology is meaningless although might have some relevance to the discussions of sloshing (provide refs).}
%\item Put this radio source's properties in the context of other BCGs (i.e. Hogan et al 2015).
%\end{itemize}

%\subsection{What shut off the radio jet?}

\section{Discussion}\label{sec:disco}
The original hypothesis we set out to address was whether a recoiling SMBH was present in this system, and thus could have helped form the large, photometrically flat, and offset core. We will first address how our new measurements affect this hypothesis.

\subsection{Assessment of knots as hypercompact stellar system candidates}
The knots were put forward by \citet{postman12} as candidates for stellar knots cloaking a recoiling SMBH. Based on the photometric and spectroscopic similarity to the BCG itself, we are now confident that these knots are in fact resident members of the \src\ core. In Section \ref{sec:knotcolors}, we demonstrated that knots 2 and 3 are consistent with being likely cannibalized %dwarf 
{small galaxies}, and given their low velocity dispersions of {168--209}\,$\kms$ are no longer a candidate HCSS for a roaming SMBH. Thus, we are left with knots 1 and 4 as HCSS candidates, however based strictly on their colors, knots 1 and 4 could feasibly also be interpreted as stripped cluster members. As previously discussed, our large error bar on knot 1 only rules out the HCSS hypothesis at a $\sim1.5\sigma$ level and thus measurements of higher S/N are still required to rule out the hypothesis confidently.

Based on our new measurements, we can attempt some consistency checks to further investigate knots 1 and 4 as potential HCSS candidates. \citet{m09} formulated a relationship between the total mass of stars bound to the black hole $M_{\rm s}$, black hole mass $M_\bullet$, host galaxy velocity dispersion $\sigma$, and initial kick velocity $V_{\rm k}$:
\begin{equation}\label{eq:vk}
\frac{V_{\rm k}}{1000\,\kms} \simeq 0.21 \bigg(\frac{\sigma}{200\,\kms}\bigg)\bigg(\frac{M_{\rm s}}{M_\bullet}\bigg)^{-2/5}~.
%\frac{M_{\rm s}}{M_\bullet} \sim 0.02 \bigg(\frac{\sigma}{200\,\kms}\bigg)^{5/2}\bigg(\frac{V_{\rm k}}{1000\,\kms}\bigg)^{-5/2}~.
\end{equation}
Through various mass estimate methods, \citet{postman12} found agreement that the central SMBH in \src\ is on the order of $M_\bullet \sim 10^{10}\,\msun$, while its velocity dispersion is $\sigma=387\,\kms$. Using the stellar mass estimates of Table \ref{table:knotphot} and the above equation, under the hypothesis that knot 1 is a stellar cloak, this formulation implies kick velocities of between 480--600\,$\kms$. For knot 4, we obtain kick velocities in the range 850--1100\,$\kms$ assuming the different IMFs.
%Other formulations relating $V_{\rm k}$ to the observed dispersion in a stellar cloak have similarly high, or higher, velocity predictions \citep[\eg][]{merrit+09}.

We would not expect the velocity offset of an observed cloak to exceed this kick velocity; it is likely to be less due to misalignment with line of sight and dampening of the SMBH velocity over time. Knot 1 appears to be well within these values despite the large error on its dynamical measurements. Should future observations permit dynamical assessment of knot 4, if it represents an HCSS we would likewise not expect its velocity offset from the core to exceed the kick velocity.

%The line-of-sight velocity offset of this object from the core is $-60\,\kms$, with an absolute value well within the predicted kick velocity, and thus this test does not provide strict evidence for or against knot 1 as a cloak.  Similarly,
We can use the HCSS size prediction of \citet{m09} to note that for kick velocities in the range 480--600\,$\kms$, the HCSS size is predicted to be up to around 200\,pc or 0\asec05 in our images, and therefore this range of velocities is consistent with the HWHM for knot 1, $\leq0.047$,  reported earlier. Thus, this comparison demonstrates consistency with an HCSS model, but not conclusive support. 
For knot 4, velocities in the range 850-1100\,$\kms$ predict smaller effective radii of between 30 and 60\,pc, or up to 0\asec015. %This knot is not resolved by present observations, and thus current information is consistent with the HCSS predictions for it. For reference knot 3, which is well-resolved, is approximately 4--5 times too large for the HCSS size prediction based on its stellar mass.

Finally, the approximate ``sloshing timescale'' of the \src\ core is $\sim10^7$\,y \citep{postman12}. We can thus note that ignoring drag effects and galaxy potential, a recoiling SMBH may have travelled 5--6\,kpc given the inferred kick velocities for knot 1, and 8.5--12\,kpc for knot 4. This is consistent with the observed offset of knot 1 from the center of both the core and the photometric center of the galaxy envelope.
Note that our support for knot 1 as an HCSS candidate differs from that of \citet{postman12} because they lacked knot mass estimates, and thus had assumed $V_{\rm k}=1000\,\kms$. 

%We can also use Eq.\,\ref{eq:vk} to 

We do not have sufficient S/N on our velocity distribution measurements to assess non-gaussianity of the spectral features, as discussed in \cite{m09}.

% NOTE We will skip the analysis of inferring Vk directly from the HCSS velocity dispersion because the error on it is too big.
% Where $\gamma$ is the power-law index of the pre-kick stellar density profile, here set at a fiducial value of $\gamma=1$.
Thus knot 1, and knot 4 (for which spectroscopy has not yet been performed), remain candidate stellar cloaks in the recoil scenario.

A2261 has two \emph{Chandra} exposures of 10 and 25 ksec (ObsIDs 550 and 5007, respectively), which do not conclusively reveal any accretion onto a SMBH.  The X-ray flux in those data is dominated by 0.5--2\,keV emission from hot cluster gas, showing a cooled core \citep{2005MNRAS.359.1481B}.  Above 2\,keV, there is some emission, possibly a point source consistent with knot 4, though it is too faint to reliably distinguish it from cluster gas emission. 

\subsection{Assessment of the radio relic in the \src\ core}
%We were interested to find that this jet has no detectable core and consists entirely of relic emission.
%The nature of the radio object was assessed previously in Sec.\,\ref{sec:radiorelic}. While it is tempting to draw ties between a jet shut-off time and a purported recoil timeframe, the evidence connecting these potential events is not yet sufficiently strong to draw solid conclusions.
The age of the relic radio source can potentially be understood as a time-marker for an event related to disturbance of the central SMBH, and thus we are interested in understanding the age and nature of this relic. We limited the timescale of last electron energization in the radio source to $\gtrsim50$\,Myr.
%, with a rapid increase in predicted age if the magnetic field in the relic is less than 50$\mu$G.
The existence of a relic in this BCG is interesting in itself and has a number of potential origins, as discussed in Sec.\,\ref{sec:radiorelic}.

To investigate how atypical \src's radio source properties are among BCGs, we compare it with the sample of \citet{hogan+15}, who systematically studied the radio properties of a large sample of BCGs (including \src). In comparison with that sample we find that the 1.4\,GHz radio luminosity is typical of $\sim$50\% of BCGs.
%\begin{itemize}
We strikingly appear to not detect a distinct core in this object down to a 3$\sigma$ limit of $<3.6\,\mu$Jy at 10\,GHz. While the non-detection of a core component in the \citet{hoganthesis} sample was relatively common, their typical upper limit on core emission was on average a factor of $>$1000 less stringent than our limit. Nevertheless, they found that $\sim$30\% of their BCGs had clear spectrally identified radio cores, while 30\% consisted of only non-core components (i.e. had upper limits on core emission, as in our case), and 30\% were complete radio non-detections; see their Table C.1.

\begin{figure}
\centering
%\vspace{3mm}
\includegraphics[trim=2.0cm 2.1cm 1.5cm 2.3cm,clip,width=0.95\columnwidth]{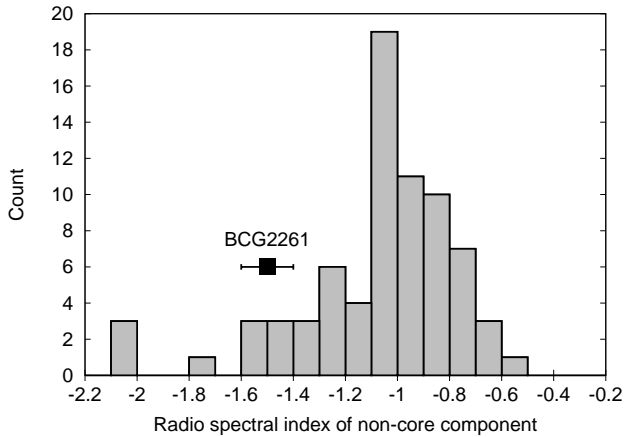}
\caption{The distribution of the spectral index for a subset of the \citet{hoganthesis} BCG sample. Here we show only the $\sim$30\% of BCGs in that sample which are similar to \src\ in that only a non-radio-core component has been spectroscopically identified in the BCG. Note that Hogan et al.'s measurement of the \src\ is included in the histogram. Our measurement is indicated separately by a square marker.}\label{fig:spixhist}
%\vspace{-1mm}
\end{figure}

The most unusual thing about the radio relic in \src\ is its central position combined with its small size and steep spectral index. Of the 30\% non-core BCGs in the Hogan sample, we show their spectral index distribution in Figure \ref{fig:spixhist}. There are only a handful of targets in the sample with spectra comparable or steeper than the central relic in \src, and it is likely that given the large selection of \citet{hoganthesis}, there is some contamination from BCG-unrelated cluster relics included in this distribution. The distribution of core-dominated BCGs tend to center on indices closer to $\alpha\sim0$. Therefore, it seems that the radio spectrum of \src\ lies at the extremity of implied activity and jet ages for BCGs.

\subsection{A Stalled Massive Perturber?}

The presumption behind our analysis of the \src\ core is that
the galaxy hosted a high-mass central BH until perhaps relatively recently.
We hypothesized that the BH has been ejected, and might be found
in one of the knots or elsewhere in and around the core.
\citet{bongra16} in contrast suggest that the core was generated
by the merger of \src\ with a ``stalled massive perturber,''
which would be the remnant central regions of another galaxy that
merged with or was cannibalised by \src.  The orbital decay of the
perturber ``stalls'' near the outer regions of a nearly
constant-density core (as \src\ has), leaving it to persist
as one of the stellar knots.

It appears that the \citet{bongra16} scenario differs from ours mainly
in whether or not \src\ and any of the galaxies that merged with
it possess central BHs.  As we noted in the introduction,
cores are generated in the standard picture as the end point of
a ``dry merger'' of two galaxies that each possess a central BH.
Ejection of the central BH in the merger occurs as one outcome
when the binary BH formed in the merger coalesces.
\citet{bongra16} do not discuss the role or even the presence
of the central BHs in their scenario,
and indeed it appears that they are not included in the theoretical models
of \citet{goerdt} on which their discussion is based.
In short, the \citet{bongra16} scenario presumes as we do that the core of A2261-BCG presently may not currently harbor a central black hole; however, they implicitly take this as an unexplained initial condition in contrast to our explicit hypothesis that the central BH has been recently ejected.
%In short, the \citet{bongra16} scenario presumes that the core of \src\ does not harbor a central black hole.
%However, the relic jet suggests that in fact the core did at one point have an active galactic nucleus, and therefore provides some evidence against this scenario.
%, but does not discuss how it arrived at this configuration. The authors further do not discuss why their model might be preferred or provide a test that discrimates it from the BH ejection model that we advocate. We cannot thus address how the present observations play into the stalled-perturber hypothesis.

\section{Summary}
To summarize, the salient properties of the \src\ are:
\begin{itemize}
\item The galaxy has a flat, 3-kpc radius core that contains at least four compact stellar knots \citep{postman12}. The knot positions, colors, and kinematics (specifically for knots 1, 2, and 3), imply that the knots are in fact residents of the BCG's core.
\item The core of \src\ is photometrically offset from the center of the surrounding envelope by $\sim 700$ pc \citep{postman12}.
\item Knots 2 and 3 are likely %dwarf 
{small} galaxies or larger stripped red sequence galaxies, as evidenced by their colors and by their {relatively} low internal stellar velocity dispersions and stellar masses.
\item Knots 1 and 4 could be dwarf galaxies, stripped cluster members, or stellar shrouds. Our photometric analysis of knot 1 demonstrates that it could be consistent with the \citet{m09} model for a hypercompact stellar system around a recoiling SMBH, with an implied kick velocity of $V_{\rm k}=480-600\,\kms$, based on our measurements. A more accurate determination of its stellar velocity dispersion
is required to test this possibility. The kick velocity estimate for knot 4 is $V_{\rm k}=850-1100\,\kms$.
\item The core hosts a radio relic, which appears to represent a past active nucleus that turned off at least 48\,Myr ago.
\end{itemize}
The first two points above provided compelling evidence that there may have been a major merger in the past which led first to a scoured core, and then to a systematic enlargement of this core due to the effects of a SMBH recoil.

With our new information about the radio relic and knots, the recoiling SMBH scenario remains a contender to explain the above properties of this object. The radio activity is roughly consistent with a core crossing timescale, while knots 1 and 4 could feasibly be consistent with the properties expected from a recoiling HCSS attached to a black hole. However, there is still yet no conclusive indicator as to the precise location of what would be a nomadic $\sim10^{10}\msun$ SMBH under this scenario.

A more mundane interpretation of this unusual galaxy's residents is that the knots are cluster members undergoing cannibalization by the BCG, the jet turned off because of a lack of accreting material, and that the core's potentially resident SMBH has no large-scale signatures to mark it. Currently, only circumstantial and theoretical links tie the global interpretation of this object more closely to the recoil hypothesis. Additional observations will be required to vet these hypotheses any further; the detection of large stellar velocity dispersions in knot 1 or 4 would be a strong indicator that either genuinely represents an HCSS.
Deeper X-ray observations could also better reveal low-level accretion onto a SMBH in the core.
% \emph{Chandra} has the angular resolution to localize a point source such that it can be determined whether it is in one of the knots or at the core photometric center.  
%For example, if the black hole has a mass of $10^{10}\,\msun$, and it is emitting in the 2--7\,kEv band at an Eddington fraction of $10^{-6}$ as a power-law with photon index $\Gamma = 1.9$, a 100 ksec observation would detect 5 photons in the band, sufficient for significant detection above the background.  
Such an observation would also be sensitive to a bow shock created by a recoiling SMBH and thus could provide strong positive evidence if it exists.

%Then some required rumination on what data would be important next on this object (better knot spectroscopy with higher S/N, for knots 1 and 4; Chandra observations; magnetic field measurements).

%\section{Conclusions}
%The radio jet does not necessarily favor a recoil, but is certainly supportive of that hypothesis. But if there was a recoil that created the offset jet that we observe, then we infer that X Y and Z must be true.
%

\section{Acknowledgements}
We thank Thomas Connor for providing early access to his Abell 2261 galaxy photometry catalog and thank Kevin Fogarty for running the {\tt iSEDfit} package on the A2261 knot photometry. This work has been supported in part by NSF award \#1458952, and in part by NASA through grants HST-GO-12065.01-A and HST-GO-14046 from the Space Telescope Science Institute, which is operated by the Association of Universities for Research in Astronomy, Inc., under NASA contract NAS 5-26555. The CLASH Multi-Cycle Treasury Program is based on observations made with the NASA/ESA \textit{Hubble Space Telescope}. 
The National Radio Astronomy Observatory is a facility of the National Science Foundation operated under cooperative agreement by Associated Universities, Inc.
Part of this research was carried out at the Jet Propulsion Laboratory, California Institute of Technology, under a contract with the National Aeronautics and Space Administration. We acknowledge a discount by GitHub that assisted with efficient collaborative development of this manuscript. {For some calculations in this paper requiring cosmological scaling, we used the online cosmology calculator \citep{cosmocalc}.}

\begin{comment}
\subsection{Other papers maybe of relevance}
Major dry mergers in BCGs\\
http://adsabs.harvard.edu/abs/2009MNRAS.396.2003L

A recent study on spectral aging in case we feel the need to (badly) estimate an age of the radio emission\\
http://adsabs.harvard.edu/abs/2013MNRAS.435.3353H

``Core sloshing,'' maybe, in a BCG leading to peculiar radio emission\\
http://www.aanda.org/articles/aa/pdf/2013/10/aa22023-13.pdf
\end{comment}

\bibliographystyle{aasjournal}
\bibliography{abel}

\end{document}